\documentclass[12pt]{article}

\input tcilatex
\begin{document}

\author{Chjan C. Lim \\
Mathematical Sciences\\
Rensselaer Polytechnic Institute\\
Troy, NY 12180\\
e-mail: limc@rpi.edu}
\title{Mean field theory and coherent structures for vortex dynamics on the plane}
\date{Submitted 16 September 1998, accepted 23 December 1998,\\
Physics of Fluids}
\maketitle

\begin{abstract}
We present a new derivation of the Onsager-Joyce-Montgomery (OJM)
equilibrium statistical theory for point vortices on the plane, using the
Bogoliubov-Feynman inequality for the free energy, Gibbs entropy function
and Landau's approximation. This formulation links the heuristic OJM theory
to the modern variational mean field theories. Landau's approximation is the
physical counterpart of a large deviation result, which states that the
maximum entropy state does not only have maximal probability measure but
overwhelmingly large measure relative to other macrostates.
\end{abstract}

\bigskip


\medskip


\newpage\ 

\section{Introduction}

The so-called mean field equation of the Onsager-Joyce-Montgomery (OJM)
theory$^{1,2,3}$ for the equilibrium vorticity distributions of a
two-dimensional inviscid and incompressible fluid (and guiding center line
charge model in plasmas) were derived more than twenty years ago but the
volume of research concerning these equations continues unabated. Recently,
the mean field thermodynamic limit for the planar vortex model has been
shown to exist by Caglioti et al $^4$, and independently by Kiessling$^5$,
and the existence of negative temperature states in this model was
demonstrated by Eyink and Spohn $^6$. Furthermore, these authors established
the maximum entropy principle$^7$ for this theory. Onsager$^1$ predicted
that the negative temperature equilibrium states exhibit a clustering of the
point vortices into coherent structures. Exact solutions of the OJM mean
field equations obtained by Chen et al $^8$ confirm the existence of such
coherent states in inviscid turbulent flows.

A OJM type equilibrium statistical theory for nearly parallel thin vortex
filaments was very recently put forth by Lions and Majda$^9$. They
established the mean field thermodynamic limit and the maximum entropy
principle for their problem. The mean field equation that Lions and Majda
obtained, is radically different from other mean field theories for vortex
dynamics in the sense that their equations are time-dependent. Nonetheless,
in a special singular limit of perfectly parallel vortex filaments, their
mean field equation reduces to the sinh-Poisson mean field equation of the
planar point vortex problem.

Concerning the relationship between the OJM and other statistical theories,
Majda and Holen$^{10}$ showed that both the OJM theory and Kraichnan's
truncated spectral theory$^{11}$ are statistically sharp with respect to the
recent infinite constraints theory of Miller$^{12,13}$ and Robert$^{14}$,
i.e., the few constraints and infinite constraints theory agree for low
energies. Chorin$^{15}$ has earlier indicated that a few constraints is
sufficient for a reasonable theory of statistical equilibrium in these
models. He also demonstrated numerically that the Miller-Robert theory is
valid only for moderate temperatures and has no Kosterlitz-Thouless
phase-transition except at zero temperature$^{16}$. Turkington and Whittaker$%
^{17}$ gave a numerical algorithm for the Miller-Robert theory. Recently,
Turkington$^{18}$ critiqued the Miller-Robert theory and presented a few
constraints equilibrium statistical theory for 2-D coherent structures,
where the equality constraints of the Miller-Robert theory are replaced by
inequalities. Turkington$^{18}$ argued that the finite ultra-violet cutoff
implicit in the Miller-Robert theory does not reflect the true inviscid
vortex dynamics as represented in the two dimensional Euler equations. In
their preprint$^{19}$, Boucher, Ellis, and Turkington presented rigorous
large deviation results and proved the existence of the mean field
thermodynamic limit for Turkington's theory.

Yet, several issues remain unresolved ---in spite of the above recent works,
it is still not completely clear where the OJM theory stands in relation to
other equilibrium statistical theories of two dimensional turbulence$%
^{11,12,13,14}$ especially with regard to their relative efficacy in
modelling real physical problems in fluid mechanics. Moreover, it is still
not clear how the OJM theory is related to the physicists' standard
variational formulations of mean field theory such as in Chapter 6 and 7 of
the text$^{20}$. It is the main aim of this paper to derive equation (\ref
{mft1}) in a new way, thereby demonstrating that the OJM theory is indeed a
mean field theory in the sense of Weiss, Peierls and Feynman$^{20}$ .
Specifically, we will prove that equation (\ref{mft1}) is the limit, as the
number of particles $N\rightarrow \infty ,$ of mean field equations that are
derived using the Bogoliubov-Feyman inequality$^{21}$ (\ref{bog}) , the
Gibbs entropy function and Landau's approximation$^{20}$. Previous
derivations of the OJM theory$^{1,2,3}$ are based on Boltzmann's entropy
function instead of Gibbs' entropy function. We note here the significant
point that our result holds for all initial vorticity distributions $q_o(x).$

A OJM-type theory for point vortex systems on a rotating sphere$^{22}$ has
recently been obtained by Lim$^{23}$, and the corresponding mean field
thermodynamic limit was established by methods similar to those of Kiessling$%
^5.$

\section{Mean field theory}

All the above statistical theories for two dimensional turbulence in
inviscid fluids have a common thread in the maximum entropy principle of
information theory$^7$. They lead to nonlinear elliptic equations, the
so-called mean field equations which have the generic form 
\[
\Delta \Psi =F(\Psi ,\gamma ) 
\]
for the stream function $\Psi $ of the mean field locally averaged vorticity
distributions $\bar{q}(x)=-\Delta \Psi ,$ and some parameters $\gamma .$
Indeed it can be shown that $\bar{q}(x)$ are exact stationary solutions of
the Euler equations. The mean field equation of the OJM theory for a single
component vortex gas is the nonlinear elliptic equation 
\begin{equation}
\Delta \Psi =ke^{-\beta \Psi }.  \label{mft1}
\end{equation}
Heuristically, this equation has the correct form for a mean field theory in
the sense that the precise interactions between particles has been replaced
by an approximate energy expression where the particles interact with some 
\textit{mean }field.

The main tool for the work in this section is the Bogoliubov-Feynman
inequality$^{20}$ for the free energy in the form proven by Feynman$^{21}$: 
\begin{equation}
F\leq F_0+\left\langle H_1\right\rangle _0\equiv F_{var},  \label{bog}
\end{equation}
where $F_0$ is the free energy based on the approximate Hamiltonian $H_0$,
where $H=H_0+H_1,$ and the second term on the right is the thermal average
of the remainder $H_1$ in the canonical measure based on $H_0.$ In other
words, 
\begin{equation}
\left\langle \cdot \right\rangle _0\equiv \frac{\int_CdV\text{ }(\cdot )%
\text{ }\exp \{-\beta H_0\}}{Z_0},  \label{zero}
\end{equation}
where the partition function $Z_0=\int_CdV$ $\exp \{-\beta H_0\}$ is based
on the approximate Hamiltonian $H_0.$ In the standard mean field theory$%
^{20} $, one choose $H_0$ so that $Z_0$ can be evaluated easily, and yet it
must have certain basic physical properties to make the theory physically
relevant. For negative temperatures, a derivation similar to Feynman's
yields 
\begin{equation}
F\geq F_0+\left\langle H_1\right\rangle _0\equiv F_{var}^{-}.  \label{bogneg}
\end{equation}

For the OJM theory of point vortices in the plane where the vortices are
identical and have vorticity or charge $\lambda $, we shall choose the
coarse-grained approximate Hamiltonian based on the division of the physical
domain $\Omega \subseteq R^2$ of area $A$ into $M$ equal boxes, 
\[
H_0=-\frac 12\sum_{i=1}^M\sum_{j\neq i}^Mn_in_j\lambda ^2\log |\vec{x}_i^0-%
\vec{x}_j^0|. 
\]
Here $n_i$ denotes the number of vortices in box $B_i$ which has area $h^2,$
and $\vec{x}_i^0$ is the location of the center of $B_i.$ Furthermore, the
total number of particles is $N$, i.e., 
\[
\sum_{i=1}^Mn_i=N. 
\]
Thus, $\vec{x}_i^0$ are no longer dependent on time but depend instead on
the lattice which is implicit in the coarse-grained Hamiltonian $H_0.$ Since
the full Hamiltonian of the point vortex model is 
\[
H=-\frac 12\sum_{i=1}^N\sum_{j\neq i}^N\lambda ^2\log |\vec{x}_i-\vec{x}_j|, 
\]
the remainder $H_1$ is given by 
\[
H_1=H-H_0 
\]
\[
=-\frac 12\sum_{i=1}^M\left( \sum_{j=1}^{n_i}\sum_{k\neq j}^{n_i}\lambda
^2\log |\vec{x}_j-\vec{x}_k|\right) 
\]
\[
-\frac 12\sum_{i=1}^M\sum_{i^{\prime }\neq 1}^M\left[ \left(
\sum_{j=1}^{n_i}\sum_{k=1}^{n_{i^{\prime }}}\lambda ^2\log |\vec{x}_j-\vec{x}%
_k|\right) -n_in_{i^{\prime }}\lambda ^2\log |\vec{x}_i^0-\vec{x}_{i^{\prime
}}^0|\right] 
\]
\medskip\ 
\[
=-\frac 12\sum_{i=1}^M\left( \sum_{j=1}^{n_i}\sum_{k\neq j}^{n_i}\lambda
^2\log |\vec{x}_j^{\prime }-\vec{x}_k^{\prime }|\right) 
\]
\[
-\frac 12\sum_{i=1}^M\sum_{i^{\prime }\neq 1}^M\left[ \left(
\sum_{j=1}^{n_i}\sum_{k=1}^{n_{i^{\prime }}}\lambda ^2\log |(\vec{x}_i^0-%
\vec{x}_{i^{\prime }}^0)+(\vec{x}_j^{\prime }-\vec{x}_k^{\prime })|\right)
-n_in_{i^{\prime }}\lambda ^2\log |\vec{x}_i^0-\vec{x}_{i^{\prime
}}^0|\right] , 
\]
where we have used 
\[
\vec{x}_j-\vec{x}_k=(\vec{x}_i^0-\vec{x}_i^0)+(\vec{x}_j^{\prime }-\vec{x}%
_k^{\prime })=\vec{x}_j^{\prime }-\vec{x}_k^{\prime }, 
\]
in the first sum, and 
\[
\vec{x}_j-\vec{x}_k=(\vec{x}_i^0-\vec{x}_{i^{\prime }}^0)+(\vec{x}_j^{\prime
}-\vec{x}_k^{\prime }) 
\]
in the second sum. In other words the primed vector $\vec{x}_j^{\prime }$
denotes the difference $\vec{x}_j-\vec{x}_i^0,$ which represents the vector
from the center of $B_i$ to vortex $j$ in it.

From 
\begin{eqnarray*}
(\vec{x}_i^0-\vec{x}_{i^{\prime }}^0)+(\vec{x}_j^{\prime }-\vec{x}_k^{\prime
}) &=&(\vec{x}_i^0-\vec{x}_{i^{\prime }}^0)+(\vec{x}_j^{\prime }-\vec{x}%
_k^{\prime }) \\
&=&(\vec{x}_i^0-\vec{x}_{i^{\prime }}^0)\left( 1+\frac{(\vec{x}_j^{\prime }-%
\vec{x}_k^{\prime })}{(\vec{x}_i^0-\vec{x}_{i^{\prime }}^0)}\right) ,
\end{eqnarray*}
one derives 
\begin{eqnarray*}
\log |(\vec{x}_i^0-\vec{x}_{i^{\prime }}^0)+(\vec{x}_j^{\prime }-\vec{x}%
_k^{\prime })| &=&\log |(\vec{x}_i^0-\vec{x}_{i^{\prime }}^0)|+\log \left( 1+%
\frac{|(\vec{x}_j^{\prime }-\vec{x}_k^{\prime })|}{|(\vec{x}_i^0-\vec{x}%
_{i^{\prime }}^0)|}\right) \\
&=&\log |(\vec{x}_i^0-\vec{x}_{i^{\prime }}^0)|+\frac{|(\vec{x}_j^{\prime }-%
\vec{x}_k^{\prime })|}{|(\vec{x}_i^0-\vec{x}_{i^{\prime }}^0)|}+O\left( 
\frac{|(\vec{x}_j^{\prime }-\vec{x}_k^{\prime })|}{|(\vec{x}_i^0-\vec{x}%
_{i^{\prime }}^0)|}\right) ^2,
\end{eqnarray*}
using the complex form of the logarithm, a step which we omit here for the
sake of brevity. Substituting this last inequality back into $H_1$ and
keeping only terms of order $|(\vec{x}_j^{\prime }-\vec{x}_k^{\prime })|/|(%
\vec{x}_i^0-\vec{x}_{i^{\prime }}^0)|,$ we get 
\[
H_1=-\frac 12\sum_{i=1}^M\left( \sum_{j=1}^{n_i}\sum_{k\neq j}^{n_i}\lambda
^2\log |\vec{x}_j^{\prime }-\vec{x}_k^{\prime }|\right) -\frac
12\sum_{i=1}^M\sum_{i^{\prime }\neq 1}^M\left(
\sum_{j=1}^{n_i}\sum_{k=1}^{n_{i^{\prime }}}\lambda ^2\frac{|(\vec{x}%
_j^{\prime }-\vec{x}_k^{\prime })|}{|(\vec{x}_i^0-\vec{x}_{i^{\prime }}^0)|}%
\right) , 
\]
where the first group of terms represent the intra-box interaction energy,
and the second group represents $O(h)$ terms in the inter-box interaction
energy.

To begin, we compute the partition function $Z_0$ which is based on the
approximate Hamiltonian $H_0,$ as follows 
\begin{eqnarray*}
Z_0 &=&\sum_sW(s)h^{2N}\exp \left( -\beta H_0\right) \\
&=&\sum_sW(s)h^{2N}\exp \left( \frac \beta 2\sum_{i=1}^M\sum_{j\neq
i}^Mn_in_j\lambda ^2\log |\vec{x}_i^0-\vec{x}_j^0|\right) ,
\end{eqnarray*}
where the summation is taken over all coarse-grained state (or macro-state) $%
s=(n_1,...,n_M)$ such that $\sum_{i=1}^Mn_i=N,$ and the degeneracy of the
macro-state $s$ is given by 
\[
W(s)\equiv \frac{N!}{n_1!...n_M!}. 
\]
The probability distribution $P_0(s)$ for a macrostate $s,$ is now defined
by: 
\begin{eqnarray}
P_0(s) &=&\int_{D^N(s)}\frac{\exp \left( -\beta H_0\right) }{Z_0}d^NA
\label{prob} \\
&=&\frac{W(s)h^{2N}\exp \left( -\beta H_0\right) }{Z_0},  \nonumber
\end{eqnarray}
where $D^N(s)$ denotes the part of phase space $D^N$ which is occupied by
the microstates associated with $s.$ Finally, we compute the free energy $%
F_0 $ as follows: 
\begin{eqnarray*}
F_0 &=&-\frac 1\beta \log Z_0 \\
&=&-\frac 1\beta \left[ N\log h^2+\log \left\{ \sum_sW(s)\exp \left( \frac
\beta 2\sum_{i=1}^M\sum_{j\neq i}^Mn_in_j\lambda ^2\log |\vec{x}_i^0-\vec{x}%
_j^0|\right) \right\} \right] .
\end{eqnarray*}

In the next step in this procedure, we will compute $\left\langle
H_1\right\rangle _0$ up to $O(h):-$%
\begin{eqnarray*}
\left\langle H_1\right\rangle _0 &=&\int\limits_{D^N}H_1(\vec{x})\text{ }P_0(%
\vec{x})\text{ }d\vec{x} \\
&=&\int\limits_{D^N}H_1(\vec{x})\frac{\exp \left( -\beta H_0(\vec{x})\right) 
}{Z_0}d\vec{x}_1...d\vec{x}_N \\
&=&\int\limits_{D^N}\left[ 
\begin{array}{c}
-\frac 12\sum\limits_{i=1}^M\left( \sum\limits_{j=1}^{n_i}\sum\limits_{k\neq
j}^{n_i}\lambda ^2\log |\vec{x}_j^{\prime }-\vec{x}_k^{\prime }|\right) \\ 
-\frac 12\sum\limits_{i=1}^M\sum\limits_{i^{\prime }\neq i}^M\left(
\sum\limits_{j=1}^{n_i}\sum\limits_{k=1}^{n_{i^{\prime }}}\lambda ^2\frac{|(%
\vec{x}_j^{\prime }-\vec{x}_k^{\prime })|}{|(\vec{x}_i^0-\vec{x}_{i^{\prime
}}^0)|}\right)
\end{array}
\right] \frac{\exp \left( -\beta H_0(\vec{x})\right) }{Z_0}d\vec{x}_1...d%
\vec{x}_N \\
&=&\int\limits_{D^N}\left[ -\frac 12\sum_{i=1}^M\left(
\sum_{j=1}^{n_i}\sum_{k\neq j}^{n_i}\lambda ^2\log |\vec{x}_j^{\prime }-\vec{%
x}_k^{\prime }|\right) \right] \frac{\exp \left( -\beta H_0(\vec{x})\right) 
}{Z_0}d\vec{x}_1...d\vec{x}_N\text{ }+O(h).
\end{eqnarray*}
where $\vec{x}\equiv (\vec{x}_1,...,\vec{x}_N)$ is a microstate$.$
Substituting the above expressions for $Z_0,$ $F_0$ into (\ref{bog}), the
free energy upper bound when the temperature $\beta >0,$ is given by 
\[
F_{var}\equiv F_0+\left\langle H_1\right\rangle _0 
\]
\[
= 
\begin{array}{c}
-\frac 1\beta N\log h^2-\frac 1\beta \log \left\{ \dsum\limits_sW(s)\exp
\left( \frac \beta 2\sum\limits_{i=1}^M\sum\limits_{j\neq i}^Mn_in_j\lambda
^2\log |\vec{x}_i^0-\vec{x}_j^0|\right) \right\} \\ 
+\dint\limits_{D^N}\left[ -\frac 12\sum\limits_{i=1}^M\left(
\sum\limits_{j=1}^{n_i}\sum\limits_{k\neq j}^{n_i}\lambda ^2\log |\vec{x}%
_j^{\prime }-\vec{x}_k^{\prime }|\right) \right] \frac{\exp \left( -\beta
H_0(\vec{x})\right) }{Z_0}d\vec{x}_1...d\vec{x}_N\text{ }
\end{array}
; 
\]
similarly for negative temperatures $\beta <0,$ the free energy lower bound
is given by 
\[
F_{var}^{-}\equiv F_0+\left\langle H_1\right\rangle _0 
\]
\[
= 
\begin{array}{c}
-\frac 1\beta N\log h^2-\frac 1\beta \log \left\{ \dsum\limits_sW(s)\exp
\left( \frac \beta 2\sum\limits_{i=1}^M\sum\limits_{j\neq i}^Mn_in_j\lambda
^2\log |\vec{x}_i^0-\vec{x}_j^0|\right) \right\} \\ 
+\dint\limits_{D^N}\left[ -\frac 12\sum\limits_{i=1}^M\left(
\sum\limits_{j=1}^{n_i}\sum\limits_{k\neq j}^{n_i}\lambda ^2\log |\vec{x}%
_j^{\prime }-\vec{x}_k^{\prime }|\right) \right] \frac{\exp \left( -\beta
H_0(\vec{x})\right) }{Z_0}d\vec{x}_1...d\vec{x}_N\text{ }
\end{array}
. 
\]
Whether the temperture is positive or negative, the integral in the above
equation can be evaluated as follows: 
\begin{equation}
\sum_sW(s)\int\limits_{B_1^{n_1}}d^{n_1}\vec{x}^{\prime }\text{ }%
....\int\limits_{B_M^{n_m}}d^{n_M}\vec{x}^{\prime }\text{ }\left[ 
\begin{array}{c}
-\frac 12\sum\limits_{i=1}^M\left( \sum\limits_{j=1}^{n_i}\sum\limits_{k\neq
j}^{n_i}\lambda ^2\log |\vec{x}_j^{\prime }-\vec{x}_k^{\prime }|\right) \\ 
\times \frac{\exp \left( \frac \beta 2\sum\limits_{i=1}^M\sum\limits_{j\neq
i}^Mn_in_j\lambda ^2\log |\vec{x}_i^0-\vec{x}_j^0|\right) }{Z_0}
\end{array}
\right]  \label{four}
\end{equation}

\[
=-\frac 1{2Z_0}\sum_sW(s)\left[ 
\begin{array}{c}
\exp \left( \frac \beta 2\sum\limits_{i=1}^M\sum\limits_{j\neq
i}^Mn_in_j\lambda ^2\log |\vec{x}_i^0-\vec{x}_j^0|\right) \\ 
\times \sum\limits_{i=1}^M\dint\limits_{B_i^{n_i}}d^{n_i}\vec{x}^{\prime }%
\text{ }\left( \sum\limits_{j=1}^{n_i}\sum\limits_{k\neq j}^{n_i}\lambda
^2\log |\vec{x}_j^{\prime }-\vec{x}_k^{\prime }|\right)
\end{array}
\right] 
\]

\[
=-\frac 1{2Z_0}\sum_s\left\{ 
\begin{array}{c}
W(s)\prod\limits_{i\neq j}^M|\vec{x}_i^0-\vec{x}_j^0|^{\frac{\beta
n_in_j\lambda ^2}2} \\ 
\times \sum\limits_{i=1}^M\dint\limits_{B_i^{n_i}}d^{n_i}\vec{x}^{\prime }%
\text{ }\left( \sum\limits_{j=1}^{n_i}\sum\limits_{k\neq j}^{n_i}\lambda
^2\log |\vec{x}_j^{\prime }-\vec{x}_k^{\prime }|\right)
\end{array}
\right\} . 
\]
An application of the Mean Value Theorem implies that the integrand in each
of the integrals in the last line above can be replaced as follows: 
\begin{equation}
\sum_{j=1}^{n_i}\sum_{k\neq j}^{n_i}\lambda ^2\log |\vec{x}_j^{\prime }-\vec{%
x}_k^{\prime }|=\frac{\lambda ^2n_i(n_i-1)}2L(n_i),  \label{L}
\end{equation}
where $L(n_i)$ is a large negative constant of the order of 
\[
\log |\vec{x}_j^{\prime }-\vec{x}_k^{\prime }|\simeq \log \frac h{\sqrt{n_i}%
}. 
\]
Therefore, the term 
\[
I=\sum\limits_{i=1}^M\dint\limits_{B_i^{n_i}}d^{n_i}\vec{x}^{\prime }\left(
\sum\limits_{j=1}^{n_i}\sum\limits_{k\neq j}^{n_i}\lambda ^2\log |\vec{x}%
_j^{\prime }-\vec{x}_k^{\prime }|\right) 
\]
is given by 
\[
I(n_i)=\sum_{i=1}^M\frac{\lambda ^2n_i(n_i-1)}2L(n_i)h^{2n_i}. 
\]
Substituting this expression in (\ref{four}), we get 
\[
\dint\limits_{D^N}\left[ -\frac 12\sum_{i=1}^M\left(
\sum_{j=1}^{n_i}\sum_{k\neq j}^{n_i}\lambda ^2\log |\vec{x}_j^{\prime }-\vec{%
x}_k^{\prime }|\right) \right] \frac{\exp \left( -\beta H_0(\vec{x})\right) 
}{Z_0}d\vec{x}_1...d\vec{x}_N 
\]
\[
\simeq \frac 1{2Z_0}\sum_s\left\{ W(s)\left( \prod_{i\neq j}^M|\vec{x}_i^0-%
\vec{x}_j^0|^{\frac{\beta n_in_j\lambda ^2}2}\right) \sum_{i=1}^M\frac{%
\lambda ^2n_i(1-n_i)}2L(n_i)h^{2n_i}\right\} . 
\]
Thus, both $F_{var}^{-}$ and $F_{var}$ up to order $O(h)$ is given by$:$%
\begin{eqnarray}
F_{var}^{-}=F_{var}=-\frac 1\beta \left[ N\log h^2+\log \left\{
\sum_sW(s)\exp \left( \frac \beta 2\sum_{i=1}^M\sum_{j\neq i}^Mn_in_j\lambda
^2\log |\vec{x}_i^0-\vec{x}_j^0|\right) \right\} \right] &&  \label{fvarr} \\
-\frac 1{4Z_0}\sum_s\left\{ W(s)\left( \prod_{i\neq j}^M|\vec{x}_i^0-\vec{x}%
_j^0|^{\frac{\beta n_in_j\lambda ^2}2}\right) \sum_{i=1}^M\lambda
^2n_i(n_i-1)L(n_i)h^{2n_i}\right\} +O(h) &&.  \nonumber
\end{eqnarray}

\subsection{Bounds}

The strategy of variational mean field theories$^{20}$ at this point is to
introduce some parameters $\gamma $ into the expressions for $F_{var}$
(respectively $F_{var}^{-})$ and minimize $F_{var}\{\gamma \}$ (resp.
maximize $F_{var}^{-}\{\gamma \})$ with respect to $\gamma .$ This yields
the best possible approximation to the full free energy $F$ within the class
determined by the chosen reduced Hamiltonian $H_0.$ Since 
\[
F=-\frac 1\beta \log Z,
\]
we now have the best approximation for $Z.$ In our derivation of the mean
field theory for the point vortex model, these parameters are given by the
macrostate $s=\{n_1,...n_M\}$ which are occupation numbers of the boxes $B_i$
in the statistical coarse-graining procedure. We will now use the facts that
the energy $H$ of the system, the entropy $S_0$ and the partition function $%
Z_0$ are all highly focussed at the equilibrium (most probable) macrostate $%
s^{*}$ in the mean field limit of large $N$. This is essentially the Landau
approximation in the Ginsburg-Landau theory for phase transitions$^{20}$. In
particular, we will use the following consequence of the Landau
approximation 
\[
P_0(s^{*})=\frac{W(s^{*})\left( \prod\limits_{j\neq i}^M|\vec{x}_i^0-\vec{x}%
_j^0|^{\frac{\beta n_in_j}2\lambda ^2}\right) }{\sum\limits_sW(s)\left(
\prod\limits_{j\neq i}^M|\vec{x}_i^0-\vec{x}_j^0|^{\frac{\beta n_in_j}%
2\lambda ^2}\right) }\simeq 1,
\]
to analyse the expressions (\ref{fvarr}). Thus $F_{var}$ and $F_{var}^{-}$ (%
\ref{fvarr}) become 
\begin{eqnarray}
F_{var}^{-}=F_{var}\simeq -\frac 1\beta \left[ N\log h^2+\log \left\{
W(s^{*})\exp \left( \frac \beta 2\sum_{i=1}^M\sum_{j\neq i}^Mn_in_j\lambda
^2\log |\vec{x}_i^0-\vec{x}_j^0|\right) \right\} \right]  &&  \label{fvarr3}
\\
-\frac 14\left\{ \sum_{i=1}^M\lambda ^2n_i(n_i-1)L(n_i)\right\}  && 
\nonumber
\end{eqnarray}
for all $\beta \geq 0$ and for a range of values $0>\beta >\beta ^{*},$
where $\beta ^{*}$ is given by the theory for the existence of the mean
field thermodynamic limit $^{4,5}.$

In order to minimize $F_{var}$ and maximize $F_{var}^{-}$ (\ref{fvarr3})
with respect to $s=\{n_1,...,n_m\}$ under the constraint 
\[
\sum_{i=1}^Mn_i=N, 
\]
we augment $F_{var}(s)$ and $F_{var}^{-}(s)$ by adding the auxillary term
with Lagrange multiplier $\alpha $ to obtain 
\begin{eqnarray*}
\tilde{F}_{var}(s) &=&F_{var}+\alpha (\sum_{i=1}^Mn_i-N), \\
\tilde{F}_{var}^{-}(s) &=&F_{var}^{-}+\alpha (\sum_{i=1}^Mn_i-N),
\end{eqnarray*}
Then taking the gradient $\nabla _s\tilde{F}_{var}(s)=0$ (resp. $\nabla _s%
\tilde{F}_{var}^{-}(s)=0)$ yields 
\[
\frac 1\beta (\log n_i+1)+\sum_{j\neq i}^M\lambda ^2n_j\log |\vec{x}_i^0-%
\vec{x}_j^0| 
\]
\begin{equation}
+\lambda ^2\left[ \frac{n_i(n_i-1)}4L^{\prime }(n_i)+\frac{2n_i-1}4L(n_i)+%
\frac{n_i(n_i-1)}2L(n_i)\right] +\alpha =0\text{ }  \label{grad1}
\end{equation}
for $i=1,...,M.$ We note that $\nabla _s\tilde{F}_{var}(s)$ (resp. $\nabla _s%
\tilde{F}_{var}^{-}(s))$ differs from $\nabla _s\tilde{F}_0(s)$ by the
derivatives of the term 
\begin{equation}
\sum_{i=1}^M\frac{\lambda ^2n_i(n_i-1)}4L(n_i),  \label{self}
\end{equation}
which is associated with the self energies of each of the $M$ boxes $B_i$ in
the coarse-graining procedure. In working with $F_{var}(s^{*})$ (resp. $%
F_{var}^{-}(s))$ instead of $F_0,$ we have put back these self energy terms.
Minimizing $F_{var}$ (resp. maximizing $F_{var}^{-}$ in the case of negative
temperatures) yields a lower bound for $F_{var}$ (resp. an upper bound for $%
F_{var}^{-})$ which provides the best approximation for the free energy $F$
of the point vortex system$^{20}$ within the ansatz of assuming $H_0$ to be
the approximate (mean field) Hamiltonian. Solving for $n_i$ in (\ref{grad1})
yields the occupation numbers 
\[
n_i=e^{-\alpha }\exp \left( -\beta \left( \sum_{j\neq i}^M\lambda ^2n_j\log |%
\vec{x}_i^0-\vec{x}_j^0|+\lambda ^2\left[ \frac{n_i(n_i-1)}4L^{\prime }(n_i)+%
\frac{2n_i^2-1}4L(n_i)\right] \right) \right) . 
\]

Next we turn to the analysis of the behaviour of this expression as $%
N,M\rightarrow \infty $ while the vortex strength is scaled by $\frac 1N,$
i.e., $\lambda =\frac 1N$ --- it is easy to see that the following limits
are valid: 
\begin{eqnarray}
\frac{n_i}N &\rightarrow &\xi (\vec{x}_i^0)\text{ }d^2x,  \label{scale1} \\
\frac{e^{-\alpha }}{Nh^2} &\rightarrow &d,  \nonumber \\
\frac 1{N^2}\left[ \frac{n_i(n_i-1)}4L^{\prime }(n_i)+\frac{2n_i^2-1}%
4L(n_i)\right] &\rightarrow &E^1(\vec{x}_i^0),  \nonumber
\end{eqnarray}
\[
\sum_{j\neq i}^M\frac{n_j}{N^2}\log |\vec{x}_i^0-\vec{x}_j^0|\rightarrow E^0(%
\vec{x}_i^0). 
\]
Thus the mean field equations in the planar point vortex theory is given by: 
\begin{equation}
\xi (\vec{x})=\Delta \Psi =d\exp \left( -\beta (E^0(\vec{x})+E^1(\vec{x}%
))\right) ,  \label{mft2}
\end{equation}
for all temperatures, which differs from that in the OJM theory by the self
energy term $E^1(\vec{x}).$

We will now analyse the self energy expression in some detail. As we let $%
N,M\rightarrow \infty $ the vortex strengths have to be scaled by $\frac 1N$
in order that the mean field (non-extensive) thermodynamic limit exists, as
was shown by Caglioti et al$^4$, Kiessling$^5$ and Eyink and Spohn$^6$.
Since $N\leq M\max_i(n_i)$ and $h^2=A/M,$ we have the bound

\begin{eqnarray*}
L(n_i) &=&\frac 12\log \frac{h^2}{n_i}\geq \frac 12\log \left[ h^2\left(
\frac 1{\max_i(n_i)}\right) \right] \\
&=&\frac 12\log A-\frac 12\log M-\frac 12\log [\max_i(n_i)] \\
&=&\frac 12\log A-\frac 12\log [M\max_i(n_i)] \\
&\geq &\frac 12\log A-\frac 12\log N,
\end{eqnarray*}
which implies that for each $i=1,...,M,$ 
\begin{equation}
|L(n_i)|\leq \frac 12\log N-\frac 12\log A.  \label{lb}
\end{equation}
The self energy term in (\ref{mft2}) scales as follows 
\begin{eqnarray*}
|E^1(\vec{x})| &\leq &\frac{n_i^2}{N^2}|L^{\prime }(n_i)|+\frac{n_i^2}{N^2}%
|L(n_i)| \\
&\leq &\frac{n_i^2}{N^2}|L^{\prime }(n_i)|+\frac{n_i\max_in_i}N\left( \frac{%
|L(n_i)|}N\right) \\
&\leq &\frac{n_i^2}{N^2}|L^{\prime }(n_i)|+n_i\frac{|L(n_i)|}N,
\end{eqnarray*}
where we have used the fact that $\left( \frac{\max_in_i}N\right) <1.$ The
first term tends to zero as $N\rightarrow \infty $ because 
\begin{eqnarray*}
\frac{n_i^2}{N^2}|L^{\prime }(n_i)| &=&\frac{n_i^2}{N^2}\left( \frac
1{2n_i}\right) =\frac{n_i}{2N^2} \\
&\leq &\frac 1{2N}\rightarrow 0.
\end{eqnarray*}
By (\ref{lb}), the second term tends to zero as $N\rightarrow \infty ,$
i.e., 
\[
n_i\frac{|L(n_i)|}N\leq \frac{n_i}N\left( -\frac 12\log A+\frac 12\log
N\right) \rightarrow 0\text{ }, 
\]
because the number $n_i$ $\sim \frac NM$ of particles in box $B_i$ stays
about the same as the total number $N$ of particles and the number $M$ of
equal boxes in the statistical coarse-graining procedure both tend to $%
\infty .$ We have shown that

\textbf{Theorem}: The mean field equation (\ref{mft2}) of the
Bogoliubov-Feynman-Landau mean field theory for point vortex dynamics tends
to the mean field equation (\ref{mft1}) of the OJM theory in the mean field
thermodynamic (non-extensive) limit of infinite particles $N\rightarrow
\infty ,$ and infinite number of boxes $M\rightarrow \infty $ in the
coarse-graining procedure in the definition of the approximate Hamiltonian $%
H_0.$

\section{Concluding remarks}

The following discussion gives a brief derivation of another formulation for 
$F_{var}$ and $F_{var}^{-}$. By definition, the free energy based on the
Hamiltonian $H_0$ is given by 
\[
F_0=\left\langle H_0\right\rangle _0-TS_0, 
\]
where the expectation operator $\left\langle \cdot \right\rangle _0$ is
defined by (\ref{zero}), and $S_0$ is the Gibbs entropy function based on $%
H_0,$ i.e., 
\[
S_0=-k_B\sum_sP_0(s)\log P_0(s), 
\]
with 
\[
P_0(s)\equiv \frac{W(s)h^{2N}\exp (-\beta H_0(s))}{Z_0}. 
\]
Thus, the upper bound $F_{var}$ (resp. lower bound $F_{var}^{-})$ for the
free energy $F$ 
\begin{equation}
F_{var}^{-}=F_{var}=\left\langle H_0+H_1\right\rangle _0-TS_0.  \label{bog2}
\end{equation}
is exactly equal to the free energy based on the full Hamiltonian $H,$ but
using the probability distribution $P_0.$

To summarize, we have derived the OJM theory by a new method which is based
on the Bogoliubov-Feynman inequality, the Gibbs entropy function and
Landau's approximation, and showed that it is a mean field theory. It is
remarkable that no additional conditions on the initial vorticity
distributions was needed to prove this result. In the case of a two
component vortex gas, our procedure gives the well-known sinh-Poisson
equation$^2$. The analogue of the above theorem for point vortices on a
rotating sphere is presented in Lim$^{23}$. We note that previous
derivations of the OJM theory$^{1,2,3}$ are based on Boltzmann's entropy
function instead of Gibbs' entropy function, which is the fore-runner of the
information-theoretic entropy function.

In another paper$^{24}$, this author will return to the issue of the
indeterminacy of the OJM mean field equations which was raised by Turkington 
\cite{Turk}. This issue concerns the fact that a given continuous vorticity
distribution can be represented in a number of different ways by clouds of
point vortices--- for example, one could use two species of vortices with
equal but opposite circulations, or one could just as well choose an
approximation based on three different species of vortices. The mean field
equations ensuing from these distinct representations of the original
continuous vorticity distributions must necessarily differ; this can be
demonstrated equally well within the traditional formulation and the current
derivation of the OJM theory.

\medskip\ 

\begin{center}
Acknowledgement
\end{center}

The author would like to thank Andy Majda for many useful discussions on
equilibrium statistics and for arranging office space at CIMS in the summer
of 1998, during which period, this paper was completed. He would also like
to thank John Chu for friendly advice over the past few years.

\end{document}